# THE CONCENTRATION OF ARTIFICIAL INTELLIGENCE IN BIG TECH AND ITS IMPLICATIONS FOR HUMAN RIGHTS IN THE EUROPEAN UNION

**Marcin Marciniak**
Institute of Theoretical Physics and Astrophysics
Faculty of Mathematics, Physics and Informatics
University of Gdańsk
Gdańsk, Poland
Email: marcin.marciniak@ug.edu.pl
ORCID: https://orcid.org/0000-0002-4176-8855

## SUMMARY

The development of advanced artificial intelligence is increasingly concentrated in a small group of vertically integrated technology companies. These firms control combinations of computing infrastructure, cloud services, data, foundation models, software ecosystems, and channels of distribution. This article argues that such concentration transforms market power into a form of private governance over the conditions in which fundamental rights are exercised. Drawing on an interdisciplinary narrative review, it examines how concentration affects equality of access, privacy, freedom of science, working conditions, access to essential resources, environmental protection, and the availability of effective remedies. Evidence from data-centre expansion in Ireland, the United Kingdom, the United States, and Chile, together with studies of automation and collective bargaining, shows that benefits may be widely distributed while decision-making power and economic returns remain concentrated. Concentration does not itself establish a rights violation, but it increases dependency, opacity, switching costs, and the risk that social and environmental costs are externalised. The article concludes that competition policy must be combined with rights-based impact assessment, interoperability, public research capacity, worker participation, and enforceable access to review and remedy.

## INTRODUCTION

Artificial intelligence is frequently described as a general-purpose technology whose effects will spread across production, public administration, science, health care, education, and everyday communication. Yet the capacity to develop and operate the most advanced systems is not evenly distributed. Training and serving frontier models require specialised processors, large-scale cloud infrastructure, extensive data, highly qualified labour, electricity, water, and access to global distribution channels. The rapid growth of these requirements favours organisations that already control several layers of the digital economy. UNCTAD and the Stanford AI Index document both the increasing cost of leading models and the growing importance of private investment in advanced AI development (Stanford Institute for Human-Centered Artificial Intelligence, 2025; UNCTAD, 2025).

This development changes the relevant policy question. The problem is no longer only whether individual AI systems are accurate, safe, or economically beneficial. It is also whether a small number of firms should control the infrastructure through which AI is developed, distributed, and incorporated into social institutions. Microsoft, Alphabet, Amazon, Meta, and other large technology companies participate in different combinations of cloud computing, model development, data collection, software, advertising, devices, and distribution. NVIDIA occupies a distinct but equally strategic position in advanced accelerators and the associated software ecosystem. The category 'Big Tech' is therefore used here functionally rather than as a fixed list: it refers to firms whose control of multiple complementary assets gives them durable influence over the direction and conditions of AI development.

Concentration is not in itself a violation of human rights. Large firms can mobilise resources, accelerate innovation, provide reliable services, and make sophisticated tools widely available. The rights-based concern arises when economic concentration creates dependency without corresponding transparency, contestability, or public accountability. A cloud provider may determine the practical availability of computing resources; a model provider may define acceptable uses and technical interfaces; a platform may control access to users; and an employer may adopt a proprietary system that workers cannot meaningfully inspect or challenge. Market power can thereby become a form of private governance.

The Charter of Fundamental Rights of the European Union provides the normative framework for this analysis. Relevant provisions concern human dignity, private and family life, personal data, freedom of expression and information, freedom of the sciences, education, non-discrimination, workers' rights, social security, health, access to services of general economic interest, environmental protection, and effective judicial protection (European Union, 2012). The Charter binds EU institutions and Member States when they implement EU law; it does not convert every private business decision into a direct Charter claim. Private conduct is nevertheless shaped by instruments such as the General Data Protection Regulation, the Digital Markets Act, the Artificial Intelligence Act, labour law, consumer law, and national rules implementing EU obligations. The AI Act is particularly important because it expressly connects the governance of high-risk AI with the protection of fundamental rights and, for specified deployers, requires a fundamental-rights impact assessment (European Parliament & Council, 2024a).

The article asks how the concentration of computing infrastructure, data, models, and distribution in Big Tech affects the practical enjoyment and protection of fundamental rights in the European Union. It advances three claims. First, concentration should be analysed across the entire technical and economic system rather than at the level of model ownership alone. Second, its human-rights effects arise through several mechanisms: unequal access, informational asymmetry, externalisation of infrastructure costs, workplace control, and weakened capacity to obtain review or change providers. Third, competition enforcement, although necessary, is not sufficient. Rights-based safeguards require public technical capacity, transparency, interoperability, worker participation, environmental accounting, and remedies that remain effective even when users depend on a dominant provider.

## MATERIALS AND METHODS

The article uses an interdisciplinary narrative-review method. It combines peer-reviewed research in economics, labour studies, environmental analysis, technology policy, and robotics with reports and legal materials from EU institutions, international organisations, national regulators, public authorities, trade unions, and companies. Sources were selected when they provided empirical evidence, documented institutional arrangements, or clarified a mechanism linking concentrated technological power to the exercise of fundamental rights. The main period of interest is 2018-2026, supplemented by earlier work on technological investment cycles, robot adoption, and the normative foundations of EU rights.

The analysis proceeds in four stages. It first identifies the assets that generate concentration: advanced compute, cloud infrastructure, data, models, software ecosystems, distribution, and supply-chain chokepoints. Secondly, it traces how control over these assets affects identifiable groups, including workers, researchers, small firms, public institutions, consumers, and communities hosting data infrastructure. Third, these effects are mapped onto rights and principles recognised in the EU legal order. Finally, existing regulatory, collective, and organisational responses are evaluated according to whether they reduce dependency, improve transparency, enable participation, or provide an effective remedy.

This is not a systematic meta-analysis and does not attempt to calculate a single quantitative measure of the human-rights impact of concentration. The examples differ in jurisdiction, technology, and evidential strength. They are used to test causal mechanisms and institutional responses, not to claim that every data centre, model, or automated workplace produces identical outcomes. The distinction between a risk to the practical enjoyment of a right and a legally established violation is maintained throughout.

## 1 THE SOURCES OF CONCENTRATED AI POWER

### 1.1 Compute capital and the research divide

The most immediate source of concentration is the scale of capital required to train and operate advanced models. Expenditure includes accelerators, high-bandwidth memory, networking, data-centre capacity, electricity, cooling, engineering labour, and repeated experimentation. The resulting fixed costs favour firms that can finance large projects internally or reserve scarce capacity years in advance. They also create a cumulative advantage: access to compute enables better models, better models attract customers and developers, and the resulting revenue finances the next generation of infrastructure.

This dynamic affects freedom of science and equality of opportunity. Research groups without industrial-scale compute may be able to study efficient methods, smaller models, or downstream applications but may be unable to reproduce, scrutinise, or challenge claims about frontier systems. Ahmed and Wahed (2020) describe a de-democratisation of AI research, while Besiroglu et al. (2024) show how unequal access to compute threatens academic contribution and independent scrutiny. Formal academic freedom may therefore coexist with a material inability to investigate the systems that increasingly shape public policy and economic life. The issue is not a general entitlement to unlimited computing resources. It is whether universities and public-interest researchers retain sufficient capacity to test claims, identify risks, and develop alternatives without dependence on the firms being evaluated.

### 1.2 Vertical integration and ecosystem control

Concentration becomes more consequential when control extends across complementary layers. A company may provide cloud infrastructure, develop foundation models, integrate them into office software or search, distribute them through an operating system, and gather usage data that improves subsequent products. Vertical integration can reduce transaction costs and deliver a coherent service, but it can also make switching difficult. A customer that has adapted data, workflows, security processes, and staff skills to one ecosystem may face substantial technical and organisational costs when moving to another provider.

The European Commission's analysis of competition in generative AI emphasises access to key inputs, partnerships, cloud services, and distribution as potential sources of market power (European Commission, 2024). The relevant unit of analysis is consequently not a single model market. It is the network of dependencies connecting chips, cloud contracts, application interfaces, developer tools, proprietary data, and access to users. Even where several models are nominally available, they may depend on the same cloud provider, accelerator supplier, or distribution platform. Apparent product variety can therefore coexist with concentrated infrastructural power.

### 1.3 From market power to private governance

A firm exercises private governance when its contractual and technical decisions determine conditions that are socially significant but difficult for affected parties to negotiate. Terms of service can define permitted research, data retention, acceptable content, model access, or the suspension of an account. Technical standards can determine which applications interoperate. Proprietary evaluation procedures can shape whether an automated decision is intelligible. These decisions remain private in legal form, yet their effects may resemble rules governing access to infrastructure or public space.

The human-rights risk is greatest when four conditions coincide: a service is important to participation in economic or social life; the provider occupies a difficult-to-replace position; decision-making is opaque; and effective review is unavailable. This framework avoids treating corporate size as intrinsically wrongful. It focuses instead on dependency, the distribution of decision-making power, and the ability of affected persons to obtain an explanation, negotiate conditions, change providers, or seek redress. These criteria also explain why concentration can affect rights even when a particular AI output is technically accurate.

## 2 ECONOMIC POWER PRODUCTIVITY AND DISTRIBUTIVE RIGHTS

### 2.1 Investment opportunity costs and public priorities

AI investment uses scarce capital, engineering capacity, electricity infrastructure, land, and public subsidies. The existence of an opportunity cost does not prove that investment in AI crowds out more valuable projects. It does require public authorities to compare expected benefits with alternative uses and to disclose how risks are allocated. This becomes a rights-based question when investment decisions affect education, health, social protection, or access to essential services. Public support for private infrastructure should therefore be assessed not only by gross investment or promised employment, but also by additionality, local benefits, fiscal risk, and the distribution of long-term infrastructure costs.

Universities illustrate the dilemma. Cloud credits and industrial partnerships can give researchers access to tools that would otherwise be unaffordable. At the same time, dependence on a small group of providers may steer research towards topics compatible with their platforms, reduce the reproducibility of results, and divert institutional resources from teaching, student support, or less commercially attractive disciplines. Freedom of science under Article 13 of the Charter and the right to education under Article 14 do not prescribe a particular university budget. They do, however, support scrutiny of whether public institutions preserve pluralism, equitable access, and independent research capacity (European Union, 2012).

Rapid investment also creates financial risk. The late-1990s telecommunications and internet boom demonstrates that overvaluation can coexist with infrastructure that remains valuable after a correction (Doms, 2004; Perez, 2002). The relevant lesson is not that current AI investment must produce the same market outcome. It is that private incentives may produce both useful capacity and duplication, leverage, or speculative pricing. Human rights enter this comparison when public guarantees, pension assets, tax concessions, or regulated utility investments transfer downside risks to households while upside returns remain concentrated.

### 2.2 Productivity without automatic social progress

Experimental and firm-level studies show that generative AI can raise worker performance, accelerate selected tasks, and support product innovation (Babina et al., 2024; Brynjolfsson et al., 2025; Noy & Zhang, 2023). These findings are economically important but do not determine how gains are distributed. Higher output per worker can lead to higher wages, lower prices, shorter working time, increased profit, or reduced employment. The outcome depends on product-market competition, ownership, labour demand, taxation, collective bargaining, and the ability of workers to influence deployment.

Concentration affects each of these channels. Firms with market power may retain a larger share of efficiency gains rather than passing them to consumers. Control of proprietary systems can make smaller firms dependent on licensing terms that limit their own margins.

Workers may become dependent on tools that increase measurable output while simultaneously enabling closer surveillance and more demanding targets. Models of automation show that productivity growth can coexist with increasing income and wealth inequality when ownership is concentrated and new human tasks are not created rapidly enough (Moll et al., 2022).

The rights-based issue is therefore not a claimed right to a specific share of every technological gain. It is the institutional capacity to participate in decisions that materially change working conditions and to protect dignity, equality, health, and economic security during adjustment. Articles 27, 28, and 31 of the Charter connect information, consultation, collective bargaining, and fair working conditions. They provide a basis for asking whether deployment remains a unilateral managerial decision or becomes a negotiated organisational change.

## 3 INFRASTRUCTURAL POWER AND ENVIRONMENTAL JUSTICE

### 3.1 Electricity networks housing and public cost

The cloud is a physical infrastructure. Large-scale AI depends on data centres connected to electricity networks, water systems, fibre networks, and transport and construction supply chains. When demand grows faster than local infrastructure, private investment decisions affect other users. The relevant effects include delays in new connections, congestion in transmission and distribution networks, higher capital expenditure, pressure on tariffs, and disputes over who should finance reinforcement. These effects transform data-centre location from a private planning matter into a question about access to services of general economic interest.

Ireland provides a clear example. The Commission for Regulation of Utilities adopted a new connection policy for data centres in response to the security and system effects of rapidly growing demand (Commission for Regulation of Utilities, 2025). In London, constrained electricity capacity has been associated with delays to housing development in areas where data-centre demand competes for grid access (Greater London Authority, 2025). Ohio has introduced data-centre-specific tariff arrangements intended to allocate infrastructure risk more directly to large users (AEP Ohio, 2025; Public Utilities Commission of Ohio, 2025). Virginia's public review similarly identified substantial implications for generation, transmission, land use, and ratepayers (Joint Legislative Audit and Review Commission, 2024).

These cases do not show that households possess an unconditional right to priority over every industrial user. They do show why allocation rules require transparency and justification. Article 36 of the Charter recognises access to services of general economic interest in accordance with national laws and practices. Where a small number of hyperscale customers can shape investment plans, public authorities should assess whether tariffs, connection agreements, and public subsidies place a disproportionate burden on households, small firms, housing, transport electrification, or heating. Concentration matters because a limited number of buyers may be large enough to affect system planning while retaining the ability to choose among jurisdictions.

### 3.2 Carbon water and territorial inequality

The environmental impact of AI cannot be evaluated only through average energy efficiency. More efficient chips and models can reduce resource use per task while lower costs expand the number and complexity of tasks. The International Energy Agency projects strong growth in electricity demand from data centres while also identifying applications through which AI could support emissions reductions in other sectors (International Energy Agency, 2025). The net effect depends on the carbon intensity of marginal electricity, the timing and location of workloads, additional renewable generation, network constraints, and rebound effects (Kaack et al., 2022; OECD, 2022).

Georgia illustrates the tension between technological investment and fossil-fuel dependence. Utility planning has linked data-centre growth to additional generation and to the continued availability of coal and gas capacity (Georgia Power, 2025; Walton, 2025). Even where a company matches annual consumption with renewable procurement, local system effects may differ by hour and location. A rights-based analysis should therefore consider health, local air pollution, affordability, and the territorial distribution of environmental burdens, not only a global corporate carbon claim.

Water conflicts in Chile provide an even more direct illustration. Data-centre cooling and associated electricity production can create significant demand in water-stressed regions. Research on Santiago describes conflicts between data-centre development, local hydrology, and competing social uses, while Chile's Second Environmental Court required climate effects to be incorporated into the assessment of the Cerrillos project (Second Environmental Court of Chile, 2024; Tironi & Albornoz, 2025). The human-rights significance lies in environmental justice: globally distributed digital benefits may be supported by locally concentrated costs borne by residents with limited influence over infrastructure decisions.

Article 35 of the Charter concerns health protection, while Article 37 requires a high level of environmental protection to be integrated into EU policies. These provisions do not establish that every increase in energy or water use is unlawful. They support lifecycle assessment, cumulative-impact analysis, public participation, and location-specific safeguards. Operators that benefit from scarce network or water capacity can reasonably be required to disclose resource use, contribute to infrastructure costs, provide demand flexibility, and show that claimed environmental benefits are additional rather than merely contractual.

## 4 WORKPLACE POWER PRIVACY AND HUMAN DIGNITY

### 4.1 Automation and unequal adjustment

AI affects work through task assistance, automation, monitoring, and organisational redesign. The distinction between automating tasks and eliminating occupations remains essential. Most jobs combine routine activities with judgement, communication, exception handling, responsibility, and tacit knowledge. A system can remove selected tasks while preserving employment, create new supervisory work, or intensify the remaining human activities. Aggregate employment may therefore remain stable even when particular workers, regions, or entry-level career paths experience substantial losses.

Evidence from industrial robotics demonstrates this diversity. Acemoglu and Restrepo (2020) find employment and wage losses in exposed US local labour markets. Dauth et al. (2021) identify displacement from German manufacturing but no comparable decline in total local employment because adjustment occurred through services and reduced entry of younger workers. Adachi et al. (2024) find that lower robot prices in Japan increased both robot use and employment by expanding output. These results caution against a universal job-loss coefficient, but they also show that national averages can conceal concentrated harm.

Big Tech concentration enters this process through the provision of general-purpose models, cloud analytics, and workplace platforms. A relatively small group of providers can standardise how performance is measured and how automated recommendations are integrated into recruitment, scheduling, promotion, or dismissal. The immediate employer remains responsible for deployment, but proprietary dependence can limit what the employer, worker representatives, or regulators can inspect. The ability to contest a decision is weakened when relevant data, model documentation, and technical expertise are distributed across several contractual layers.

### 4.2 Monitoring safety and autonomy

AI-supported systems can remove dangerous tasks, reduce heavy lifting, and improve predictive maintenance. Gihleb et al. (2022) associate robot exposure with lower injury rates, particularly in manufacturing. Yet automation can also accelerate the pace of remaining work. Burtch et al. (2025) find that warehouse robotics reduced severe injuries while increasing less severe injuries, illustrating how safety gains in one category may coexist with work intensification in another. Research across European countries also links robotisation with reduced perceived meaningfulness and autonomy in some settings (Nikolova et al., 2024).

Continuous data collection creates a direct connection with privacy and personal-data protection. Workplace systems may record location, keystrokes, voice, images, productivity, error rates, emotional indicators, and interaction patterns. Even where each data point appears innocuous, combining them can produce detailed profiles and influence access to employment. The GDPR requires a lawful basis, purpose limitation, data minimisation, and safeguards for data subjects, while the Platform Work Directive introduces specific protections concerning automated monitoring, automated decision-making, human oversight, and review in platform work (European Parliament & Council, 2016, 2024b).

Human dignity is relevant because workers should not be reduced to continuously optimised data points. Fair working conditions require attention to pace, safety, predictability, and the possibility of meaningful human intervention. Non-discrimination requires testing outcomes across age, gender, disability, migration status, and other protected characteristics. Effective remedy requires more than a generic statement that AI was used. Affected persons need intelligible reasons, access to relevant evidence, a competent human reviewer, and protection against retaliation for challenging a system.

### 4.3 Care robotics and essential human contact

Care robots show why the same technology can support and undermine rights depending on institutional choices. Robots can assist mobility, rehabilitation, reminders, monitoring, and communication, potentially strengthening the independence of older persons and persons with disabilities. Evidence suggests that social robots can reduce loneliness in some settings, particularly when they supplement rather than replace human support (Mehrabi & Ghezelbash, 2025).

The risk arises when automation is used primarily to reduce staffing. Assistance may become technically continuous but socially thinner. Concerns include loss of human contact, infantilisation, deception about a machine's emotional capacities, intrusive monitoring, and ambiguous consent among persons with cognitive impairment (Sharkey & Sharkey, 2012). When the robot, model, cloud service, and care platform come from different providers, responsibility can become fragmented. Rights to dignity, integrity, privacy, health, and an effective remedy then depend on clear procurement rules and an identifiable institution that remains accountable for the complete service.

## 5 SEMICONDUCTOR CLOUD AND GEOPOLITICAL DEPENDENCE

The physical infrastructure of AI is produced through a network linking chip architecture, design software, manufacturing equipment, materials, fabrication, high-bandwidth memory, advanced packaging, server assembly, networking, and cloud operation. Concentration occurs at several non-interchangeable stages. ASML is uniquely important in extreme-ultraviolet lithography; TSMC is central to advanced foundry production and packaging; a small group of Korean and US firms supplies high-bandwidth memory; NVIDIA combines leading accelerators with a widely adopted software ecosystem; and a small number of hyperscalers operate global cloud platforms. OECD analysis emphasises that aggregate market shares can understate dependence on specific, difficult-to-substitute inputs (OECD, 2023, 2025).

These dependencies create both corporate and geopolitical power. Governments can restrict exports of advanced processors, equipment, design tools, or cloud services. Firms can allocate scarce capacity, set licensing terms, change interfaces, or decline to serve particular customers and jurisdictions. At the same time, no actor is fully autonomous: US design and cloud firms depend on Asian fabrication, memory, and assembly; foundries depend on foreign equipment and materials; and equipment suppliers rely on specialised international components. The result is asymmetric interdependence rather than national or corporate self-sufficiency.

Digital sovereignty should therefore not be equated with autarky. From a rights perspective, it is the practical capacity of public institutions to protect data, audit systems, maintain essential services during disruption, change suppliers, and negotiate acceptable conditions. A state may retain formal authority while lacking the technical capacity to verify systems used in health care, education, policing, or social administration. This weakens its ability to guarantee rights and to provide an effective remedy. Supplier diversification, open standards, portable data and models, public-interest compute, and reciprocal arrangements with trusted partners are consequently rights-enabling measures as well as industrial policy.

## 6 COUNTERVAILING POWER AND RIGHTS BASED GOVERNANCE

### 6.1 Competition interoperability and public capacity

Competition policy addresses exclusionary conduct, self-preferencing, tying, and barriers to market entry. The Digital Markets Act adds ex ante obligations for designated gatekeepers and reflects the judgment that some forms of platform power cannot be addressed only after harm has occurred (European Parliament & Council, 2022). In AI, however, competition policy must consider vertical dependencies and technical switching costs. The existence of several applications does not guarantee contestability if they depend on the same cloud, accelerator, model, or identity infrastructure.

Interoperability and portability can reduce dependency, but they require careful design. Data portability without compatible formats or affordable transfer is largely formal. Model portability may be constrained by proprietary interfaces, safety controls, hardware optimisation, or licence terms. Public procurement can create leverage by requiring documented interfaces, exportable logs, audit access, continuity plans, and credible exit arrangements. Public or shared compute facilities can support universities, small firms, regulators, and civil-society research, provided that access is allocated transparently and does not reproduce existing inequalities.

### 6.2 Fundamental rights assessment and effective remedy

The AI Act connects technical risk management with fundamental-rights protection. Article 27 requires specified deployers of certain high-risk systems to assess effects on fundamental rights before use (European Parliament & Council, 2024a). Its direct scope is narrower than the full political economy examined here. It does not turn every cloud contract, data-centre investment, or model partnership into a mandatory fundamental-rights assessment. It nevertheless provides a useful institutional model: identify affected groups, describe foreseeable impacts, specify oversight and mitigation, and reconsider deployment when risks cannot be adequately controlled.

Assessment should not become a one-off compliance document. Before deployment, organisations should identify the purpose, alternatives, affected groups, data sources, dependencies, and exit options. During use, they should track errors, discriminatory outcomes, workload, resource consumption, security incidents, and complaints. After a material change in the model, provider, or context, the assessment should be updated. Independent regulators, worker representatives, and affected communities require access to enough information to test whether safeguards operate in practice.

### 6.3 Collective bargaining and worker participation

Collective bargaining can convert abstract rights into enforceable workplace rules. The 2023 Writers Guild of America agreement established that AI-generated text could not be used to reduce a writer's credit or compensation and that writers could not be required to use AI. SAG-AFTRA negotiated consent and compensation safeguards for digital replicas. NewsGuild agreements have addressed protection of bargaining-unit work and human oversight of AI-assisted journalism (NewsGuild-CWA, 2025; SAG-AFTRA, 2023; Writers Guild of America, 2023). These examples show that authorship, voice, likeness, and professional judgement can remain subjects of negotiation rather than being treated as data available for unilateral reuse.

European experience provides additional models. Eurofound (2025) documents agreements and social dialogue on AI across several sectors, although coverage remains uneven. Spain's Riders Law requires information for worker representatives about parameters and rules underlying algorithms that influence working conditions and employment (Spain, 2021). The Platform Work Directive establishes rules on data processing, information, human oversight, review, and consultation for digital labour platforms (European Parliament & Council, 2024b). These measures are important because individual workers rarely possess the technical expertise or bargaining power required to challenge a complex system alone.

### 6.4 Transition policy and shared gains

Training is necessary but insufficient. It is effective only when suitable jobs exist, workers can participate during paid time or with income support, qualifications are recognised, and programmes correspond to actual labour demand. Joint programmes can improve these conditions. The Microsoft-AFL-CIO partnership links education with worker input into AI development, while Singapore's Company Training Committee model ties public support to transformation plans and measurable worker outcomes (Microsoft & AFL-CIO, 2023; National Trades Union Congress, 2025). Germany's transformation hubs provide advisory capacity for firms and workers that cannot maintain their own specialist teams (German Federal Ministry of Labour and Social Affairs, 2023).

These programmes should be evaluated by placement, wage progression, job retention, and access across age, gender, disability, employment status, and region, rather than by enrolment alone. Where productivity gains are measurable, shorter working time, gain-sharing, profit-sharing, or transition funds can distribute benefits more broadly. Where redeployment fails, unemployment protection, wage insurance, mobility assistance, pension-credit protection, and targeted regional investment remain necessary. Countervailing power requires both universal minimum standards and institutions capable of negotiating above that floor.

## CONCLUSIONS

The concentration of AI development in Big Tech is not merely a competition issue. Control over compute, cloud services, data, foundation models, software ecosystems, and distribution gives a small number of companies the ability to shape the material and institutional conditions in which AI is used. This power can support innovation and make advanced capabilities widely available. It can also create dependencies that are difficult for workers, researchers, smaller firms, public institutions, and governments to escape.

The human-rights consequences arise through identifiable mechanisms. Unequal access to compute can weaken freedom of science and reinforce economic inequality. Control of data and proprietary models can affect privacy, non-discrimination, and access to an effective remedy. Data-centre expansion can transfer electricity, water, and infrastructure costs to local communities. Workplace AI can improve productivity and safety while enabling surveillance, work intensification, and unilateral organisational control. Dependence on semiconductor and cloud chokepoints can constrain the capacity of states to maintain essential services and enforce their own rules.

None of these mechanisms establishes that large technology companies inevitably violate rights. The decisive questions concern accountability, alternatives, participation, and the allocation of costs. Concentrated power is most problematic when affected persons cannot understand a decision, negotiate conditions, change providers, or obtain review. A rights-based AI policy should therefore treat contestability as a practical institutional condition, not merely the theoretical presence of another product in the market.

The European Union already possesses elements of such an approach in the Charter, GDPR, Digital Markets Act, AI Act, Platform Work Directive, competition law, and environmental regulation. The remaining challenge is to connect them. Competition, data protection, labour rights, environmental planning, public procurement, and industrial policy should be evaluated as parts of one governance system. The aim is not to prevent scale where scale produces real benefits. It is to ensure that technological scale does not become unaccountable power over the exercise of fundamental rights.

## RECOMMENDATIONS

First, EU and national competition authorities should assess concentration across the full AI stack. Reviews of partnerships and acquisitions should consider compute reservations, cloud dependence, model access, distribution, data advantages, software ecosystems, and switching costs, rather than treating each layer as a separate market.

Second, public procurement of AI and cloud services should require interoperability, data and log portability, documented interfaces, independent audit access, incident reporting, continuity arrangements, and a credible exit plan. Public institutions should not adopt systems that they cannot meaningfully supervise or replace.

Third, fundamental-rights impact assessment should be used beyond its minimum legal scope when AI affects access to employment, education, health care, public benefits, essential services, or information. Assessments should identify affected groups, alternatives, cumulative effects, provider dependencies, and mechanisms for human review and remedy.

Fourth, the EU and Member States should expand public-interest computing capacity for universities, regulators, small firms, and civil society. Allocation should be transparent and should support independent evaluation, reproducibility, multilingual research, and projects that do not align with the commercial priorities of dominant firms.

Fifth, data-centre connection agreements and environmental permits should allocate infrastructure costs to the users that cause them, require credible demand flexibility, disclose energy and water use, and evaluate cumulative local effects. Communities should have access to relevant information and effective participation before irreversible infrastructure decisions are made.

Sixth, workers and their representatives should receive timely information before AI deployment and should be able to negotiate monitoring, performance targets, staffing, training, redeployment, human review, and the distribution of productivity gains. Legal and contractual protections should extend as far as possible to subcontracted, temporary, platform, and freelance workers who are often outside conventional bargaining structures.

Finally, every person materially affected by an AI-supported decision should have access to an identifiable human decision-maker, intelligible reasons, relevant evidence, and an effective channel of appeal. Responsibility should not disappear among the model developer, cloud provider, systems integrator, employer, and public authority. The institution deploying the system should remain accountable for the service as a whole.